\documentstyle[emulateapj,graphicx]{article}
\def\etal  {{et~al.}\ }

\def\vol#1  {{{#1}{\rm,}\ }}
\def\lya{{\rm Ly}$\alpha$\ }

\def\etal{et al.\ }

\def\gsim{\;\rlap{\lower 2.5pt
 \hbox{$\sim$}}\raise 1.5pt\hbox{$>$}\;}
\def\lsim{\;\rlap{\lower 2.5pt
   \hbox{$\sim$}}\raise 1.5pt\hbox{$<$}\;}

\newcount\refno
\newcount\rfno
\def\eq{$^{\the\refno\ }$\advance\refno by 1}
\def\ad{\advance\rfno by 1}

\def\clock{\count0=\time \divide\count0 by 60
     \count1=\count0 \multiply\count1 by -60 \advance\count1 by \time
     \number\count0:\ifnum\count1<10{0\number\count1}\else\number\count1\fi}

\def\myputfigure#1#2#3#4#5%
{\hskip0.03\textwidth\vskip#5pt
\makebox[0pt]{\hskip#2in
\includegraphics[width=#3\textwidth]{#1}}\vskip#4pt\hfill}
\newcommand{\beq}{\begin{equation}}
\newcommand{\eeq}{\end{equation}}


\begin{document}
\title{Quasar Str\"omgren Spheres Before Cosmological Reionization}
\author{Renyue Cen and Zoltan Haiman\altaffilmark{1}}
\affil{Princeton University Observatory, Princeton University, Princeton, NJ 08544}
\altaffiltext{1}{Hubble Fellow}

\begin{abstract}

Ionizing sources embedded in the neutral intergalactic medium (IGM)
before cosmological reionization generate discrete HII regions.  We
show that a sufficiently bright quasar (for example, one tenth as
luminous as that recently discovered by Fan et al.) can ionize a large
volume, allowing the transmission of a substantial fraction of the
flux of its intrinsic \lya emission line on both sides of the \lya
wavelength.  The observed line profile is richly informative.  We show
that a sufficiently accurate, high spectral resolution ($R\approx
10^4$) measurement of the line profile of a bright quasar is feasible
using the {\it Next Generation Space Telescope (NGST)} as well as
large ground-based telescopes.  Such a measurement has two potentially
important applications.  First, the red side of the \lya emission line
provides a way to measure the quasar lifetime.  Second, the blue side
provides a direct measure of the density fluctuations of the
intergalactic medium at the quasar redshift.  The estimate of the
absorption of the red side is limited by the accuracy to which the
intrinsic profile of the \lya emission line is known.  The blue side,
however, does not sensitively depend on the intrinsic profile, because
the former is much narrower than the latter.
\end{abstract}


\section{Introduction}

The epoch of cosmological reionization, and the nature of the ionizing
sources bear fundamental cosmological importance, but are yet unknown.
In contrast, we have a better understanding of how the reionization
might have happened: individual HII regions, whether around galaxies
or quasars (or sources of other nature), start out as isolated
islands, eventually overlap to fill the universe and complete the
reionization process (Aaron \& Wingert 1972; Shapiro \& Giroux 1987;
Gnedin \& Ostriker 1997; Haiman \& Loeb 1998; Miralda-Escud\'e et
al. 1999).  In this {\it Letter} we suggest that it may be possible to
directly probe the structure of an HII region around a bright quasar
before the reionization is complete.  Although we do not necessarily
have to identify the primary reionization sources with quasars, the
existence of bright quasars at high redshift is supported by the
recent discovery of the very bright quasar at $z=5.8$ in the Sloan
Digital Sky Survey (SDSS; Fan \etal 2000; F00 hereafter).

To study the size and structure of an individual Str\"omgren sphere,
we propose to utilize the observed profile of the quasar's \lya
emission line.  When a source is embedded in a neutral IGM, its
emission line suffers from strong absorption by the damping wing of
the resonance \lya absorption of the intervening neutral IGM
(Miralda-Escud\'e 1998).  As a result, the \lya photons are absorbed
and re--emitted multiple times in the IGM, resulting in an extended,
low--surface brightness line, which is difficult to observe (Loeb
\& Rybicki 1999).  In contrast, if the source is surrounded by a
large HII region, the \lya photons can escape absorption, provided
they redshift out of resonance before they reach the boundary of the
HII region.  In this {\it Letter}, we show that a sufficiently bright
quasar placed at $z\ge 6$ can produce a sufficiently large Str\"omgren
sphere to render its \lya emission line detectable, using the {\it
Next Generation Space Telescope (NGST)}, as well as high spectral
resolution ground based telescopes such as Keck and the Hobby-Eberly
(HET) telescopes.  Throughout this {\it Letter}, we adopt a flat
$\Lambda$CDM background cosmology dominated by a cosmological constant
and cold dark matter, with density parameters
$(\Omega_\Lambda,\Omega_0,\Omega_{\rm b})=(0.7,0.26,0.04)$, and Hubble
constant $H_0=70~{\rm km~s^{-1}}$.

\section{Str\"omgren Spheres Around Quasars}

According to the classical definition in stellar astronomy, a
Str\"omgren sphere is an HII region around an individual,
ionizing-photon-emitting O/B star, within which the rate of
recombinations exactly balances the emission rate $\dot N_{ph}$ of
ionizing photons from the star.  The radius $R_s$ of the Str\"omgren
sphere is given by
\begin{equation}
R_s = 
\left[ \frac{3\dot N_{ph}}{4\pi\alpha_B\langle n_H^2 \rangle}
\right]^{1/3},
\label{eq:Rs}
\end{equation}
where $\alpha_B$ is the hydrogen recombination coefficient at
$T=10^4$K, and $\langle n_H^2 \rangle$ is the mean squared hydrogen
density within $R_s$.

An analogous situation arises on much larger scales, for cosmological
ionizing sources such as galaxies and quasars (Shapiro \& Giroux
1987).  There are, however, important differences between the stellar
and cosmological case. The most important one for our purposes is that
the hydrogen recombination time in the IGM is longer than the Hubble
time at the relevant epochs ($z\lsim 10$) and the quasar lifetime. As
a result, the size $R_s$ of the {\it equilibrium} Str\"omgren sphere
is large, but is never reached in practice before the ionizing source
turns off.  The radius $R_{t_Q}$ of the ionized region around a steady
source of age $t_Q$ is then simply
\begin{equation}
R_{t_Q} =
\left[ \frac{3\dot N_{ph} t_Q}{4\pi \langle n_H \rangle}
\right]^{1/3},
\label{eq:Rt}
\end{equation}
where $\langle n_H \rangle$ is the mean hydrogen density within
$R_{t_Q}$. A second difference is that in the stellar case, the
Str\"omgren sphere is static if the luminosity of the source is
constant, whereas in the quasar case the sphere around a long--lived
quasar can evolve due to the evolution of density around a quasar
(both mean density and clumping factor) as well as the accumulation of
ionizing photons (equation 2 and equation 4 below), even if the source
luminosity remains constant.  To fully take these effects into
account, one needs to solve the equation of motion for the ionization
front, $R_i$, with the following equation
\begin{equation}
\frac{dR_i^3}{dt}= 3H(z) R_i^3 +
 \frac{3 \dot N_{ph}}{4\pi \langle n_H \rangle}
 - C_{\rm HII} \langle n_H \rangle \alpha_B R_i^3,
\label{eq:Ri}
\end{equation}
where $H(z)$ is the Hubble constant at $z$, and $C_{\rm HII}\equiv
\langle n_H^2 \rangle / \langle n_H \rangle^2$ is the mean clumping 
factor of ionized gas within $R_i$.  The three terms on the right side
in equation~\ref{eq:Ri} account for the Hubble expansion, the
ionizations by newly produced photons, and recombinations (Shapiro \&
Giroux 1987; Haiman \& Loeb 1997), respectively.  Note that
equation~\ref{eq:Ri} predicts an initial phase of superluminal
expansion for the I--front, i.e. when $R_i(t) > ct$.  Ignoring any
clumping or recombinations, this phase would last for $t
\approx 10^7 (\dot N_{ph}/10^{57}{\rm s^{-1}})^{1/2} [(1+z)/8]^{-3/2}$
years (cf. eq.~\ref{eq:Rt}); a potentially significant effect, since
the quasar found by F00 in the SDSS survey implies an intrinsic
ionizing emissivity of $\dot N_{ph} \approx 2\times 10^{57}~{\rm
s}^{-1}$.  Nevertheless, the Ly$\alpha$ photons emitted by the quasar
lag behind the I--front during the phase of relativistic expansion,
and will only see the opacity once the front slows down to be
subluminal. As a result, for our purposes, the apparent size of the
HII region is always given by equation~\ref{eq:Ri}.

In practice, we find that the apparent size of a quasar Str\"omgren
sphere is primarily determined by the total number of ionizing photons
emitted. Assuming that the quasar luminosity $\dot N_{ph}$ is
constant, the solution of equation~\ref{eq:Ri} is accurately described
by $R_{t_Q}$ given in eq.~\ref{eq:Rt}. Gas clumping is unlikely to
significantly change this answer.  As a specific example, with the
ionizing photon emissivity of the SDSS quasar, and for $C_{\rm
HII}=100$, we find that $R_i$ would be reduced by 10\% at
$\approx10^7~{\rm yr}$, and by 50\% at $\approx10^8~{\rm yr}$.
Cosmological simulations indicate that the value of $C_{\rm HII}$ is
likely to be significantly below $100$ (Gnedin \& Ostriker 1997).

\section{Ly$\alpha$ Transfer Across a Str\"omgren Sphere}

The main purpose of this {\it Letter} is to assess the effect of the
quasar's HII region on the observed profile of its Ly$\alpha$ line.
Consider a source located at $z_s\gsim 6$ before reionization is
complete, whose age is $t_Q$.  The source is assumed to be surrounded
by a spherical HII region of size $R_i(t_Q)$, but is otherwise
embedded in a neutral IGM.  The source is also assumed to emit a
Ly$\alpha$ emission line, which is subsequently reprocessed by the
opacity of the intervening neutral IGM, as well as of the residual
neutral hydrogen within the HII region itself.

The optical depth between the source and the observer at $z=0$, at the
observed wavelength $\lambda_{\rm obs}=\lambda_s(1+z_s)$, is given by
\begin{equation}
\tau(\lambda_{\rm obs},z_s)=\int_{z_r}^{z_s} dz c \frac{dt}{dz} n_H(z)
\sigma_\alpha[\lambda_{\rm obs}/(1+z)],
\label{eq:dtaudz}
\end{equation}
where $cdt/dz$ is the line element in the assumed $\Lambda$CDM
cosmology, $n_H$ is the neutral hydrogen density, and $\sigma_\alpha$
is the Ly$\alpha$ absorption cross--section.  The reionization
redshift is assumed conservatively to be $z_r=6$, but the precise
value (as well as the details of the reionization process) is
irrelevant here, since we are only sensitive to the optical depth near
the source redshift $z_s$, where the IGM is assumed to be still fully
neutral. It is useful to divide the range of integration in
equation~\ref{eq:dtaudz} into two parts, to reflect the contributions
to the optical depth from within ($z_i < z < z_s$) and outside ($z_r <
z < z_i$) the HII region, where $z_i$ is the redshift somewhat below
$z_s$, corresponding to the boundary of the HII region ($z_i\approx
z_s-R_i[t]/R_H[z_s]$, where $R_H[z_s]$ is the size of the cosmological
horizon at $z_s$).

The Ly$\alpha$ absorption cross--section is assumed to be described by
a numerically computed Voigt profile (see eq.~6 of Press \& Rybicki
1993). The gas temperature for this profile is taken to be $T=10^4$K
within the HII region, and $T=1.2[(1+z)/8]^{-2}~{\rm K}$ outside
it.\footnote{We assume that the IGM temperature is coupled to the
cosmic microwave background until $z=150$, and cools adiabatically
thereafter.}  Outside the HII region, the neutral hydrogen density is
assumed to follow the IGM density, $8.5\times10^{-5}
[(1+z)/8]^{3}~{\rm cm^{-3}}$.  Inside the HII region, the neutral
hydrogen density depends on the ionized fraction.  Assuming
photoionization equilibrium (justified for the source lifetimes of
$>10^6$ yr considered here), and that the gas is highly ionized, the
neutral hydrogen fraction $x=n_{HI}/n_H\ll 1$ at the quasar-centric
radius $r$ is given by
\begin{equation}
\alpha_B C_{\rm HII} \langle n_H \rangle = x \int_{\nu_H}^{\infty} d\nu
\frac {L_\nu}{4\pi r^2} \frac{\sigma_\nu}{h\nu} = 
\frac{x \bar\sigma\dot N_{ph}}{4\pi r^2}.
\label{eq:x1}
\end{equation}
Here $L_\nu$ is the intrinsic luminosity of the source in ${\rm
erg~s^{-1}~Hz^{-1}}$, $\nu_H$ and $\sigma_\nu$ are the ionization
threshold and cross section of hydrogen, with the luminosity weighted
cross section being
\begin{equation}
\bar \sigma\equiv
\int_{\nu_H}^{\infty} d\nu \frac {L_\nu\sigma_\nu}{h\nu} 
\times
\left[
\int_{\nu_H}^{\infty} d\nu \frac {L_\nu}{h\nu}
\right]^{-1}
\label{eq:sigeff}
\end{equation}
and equal to $2.5\times10^{-18}~{\rm cm^2}$ for the Elvis et
al. (1994) template spectrum (the same value would be obtained for a
power law spectrum with a slope of $\nu^{-1.8}$).  The resulting
neutral hydrogen fraction is
\begin{equation}
x=10^{-6} C_{\rm HII}
\left(\frac{r}{1~{\rm Mpc}}\right)^{2}
\left(\frac{\dot N_{ph}}{10^{57}~{\rm s^{-1}}}\right)^{-1}
\left(\frac{1+z}{8}\right)^{3}.
\label{eq:x2}
\end{equation}
This equation applies within the optically thin central region of the
Str\"omgren sphere where $x\ll 1$, and shows that for bright sources
of interest here, this region is highly ionized. Near the boundary of
the HII region, the neutral fraction rapidly increases to unity, where
the $x\ll 1$ condition breaks down and equation~\ref{eq:x2} becomes
invalid. However, this does not affect the results shown below for the
following reasons.  First, the flux transmission shortward of the \lya
wavelength that we are concerned with occurs in regions with optical
depth $\tau\le 7$ or $x\le 5\times 10^{-5}$. Second, the optical depth
longward of the \lya wavelength is roughly $0.5 x
(\Delta\lambda/100\AA\,)^{-1}$ for the cosmological parameters adopted
here (Miralda-Escud\'e 1998), and is dominated by neutral ($x=1$) IGM
outside the Str\"omgren sphere. Therefore, the boundary region of the
Str\"omgren sphere does not affect the results in a sensitive way. We
assume that the jump from $x(R_i)$ to unity takes place as a step
function at the radius $R_i(t_Q)$.

\myputfigure{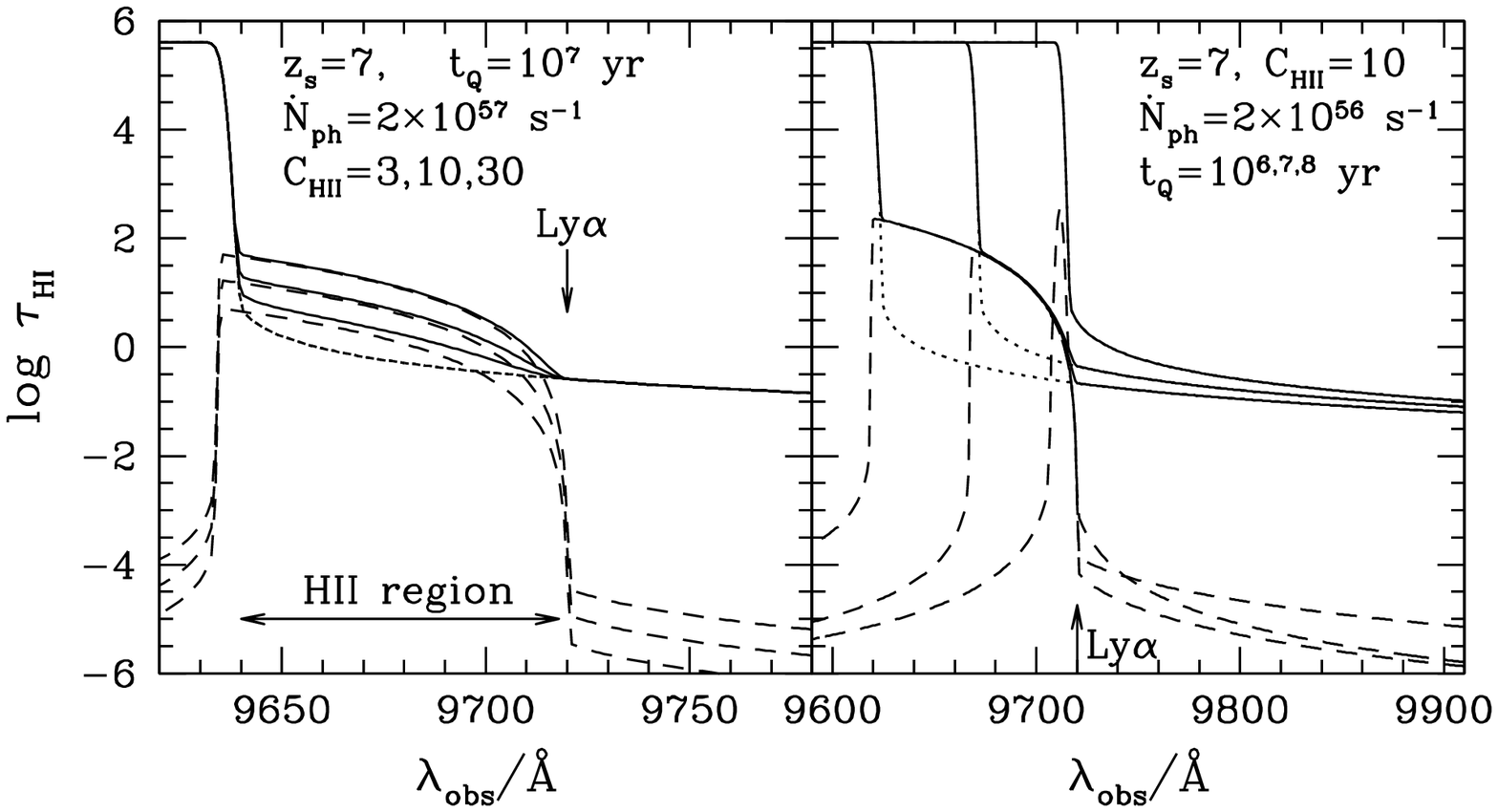}{3.2}{0.45}{-10}{-10} 
\vspace{-3.5cm} 
\figcaption{The total line optical depth (solid curves) towards
a quasar at $z_s=7$, as a function of observed wavelength. Also shown
are the separate contributions from the neutral IGM (dotted curves)
and from the residual neutral hydrogen within the HII region (dashed
curves). In the left panel, we assume an ionizing photon emissivity of
$2\times10^{57}~{\rm s^{-1}}$, a fixed quasar lifetime of $10^7$ yr,
and demonstrate the effect of the clumping factor ($C_{\rm
HII}=3,10,30$).  Clumping effects primarily the blue side of the
observed Ly$\alpha$ line. In the right panel, we assume an ionizing
photon emissivity of $2\times10^{56}~{\rm s^{-1}}$, a fixed clumping
factor of $C_{\rm HII}=10$, and demonstrate the effect of the quasar
lifetime ($10^6,10^7,10^8$ yr).  The lifetime effects primarily the
red side of the observed Ly$\alpha$ line.
\label{fig:tau}}
\vspace{\baselineskip}

\section{Results and Discussion}

Figure~\ref{fig:tau} shows the optical depth around the \lya line for
a quasar at $z_s=7$.  The left panel shows that, for a bright quasar
like that of F00, the optical depth longward of the \lya wavelength is
nearly independent of the clumping factor, simply because the optical
depth in that region is dominated by the damping wing of the neutral
IGM (dotted curves in the left panel) outside the Str\"omgren sphere,
whose radius is insensitive to $C_{\rm HII}$ as long as $C_{\rm
HII}\le 100$ (see \S 2).  The situation is dramatically different for
a fainter quasar as shown in the right panel of Figure~\ref{fig:tau};
because the HII region is much smaller, the red damping wing produced
by the neutral IGM outside the Str\"omgren sphere start to reflect the
size of the sphere hence the age of the quasar.  The dependence of the
red damping wing on the clumping factor is very weak, because the
contribution from the residual neutral hydrogen within the Str\"omgren
sphere is small.

On the other hand, the optical depth shortward of the \lya wavelength
within the stretch of the Str\"omgren sphere is always dominated by
the residual neutral hydrogen inside the Str\"omgren sphere (dashed
curves), and therefore is proportional to $C_{\rm HII}$
(equation~\ref{eq:x2}), in sharp contrast with the situation at the
longward of the \lya wavelength.  However, the optical depth shortward
of the \lya wavelength it is quite insensitive to $t_Q$
(equation~\ref{eq:x2}).

\myputfigure{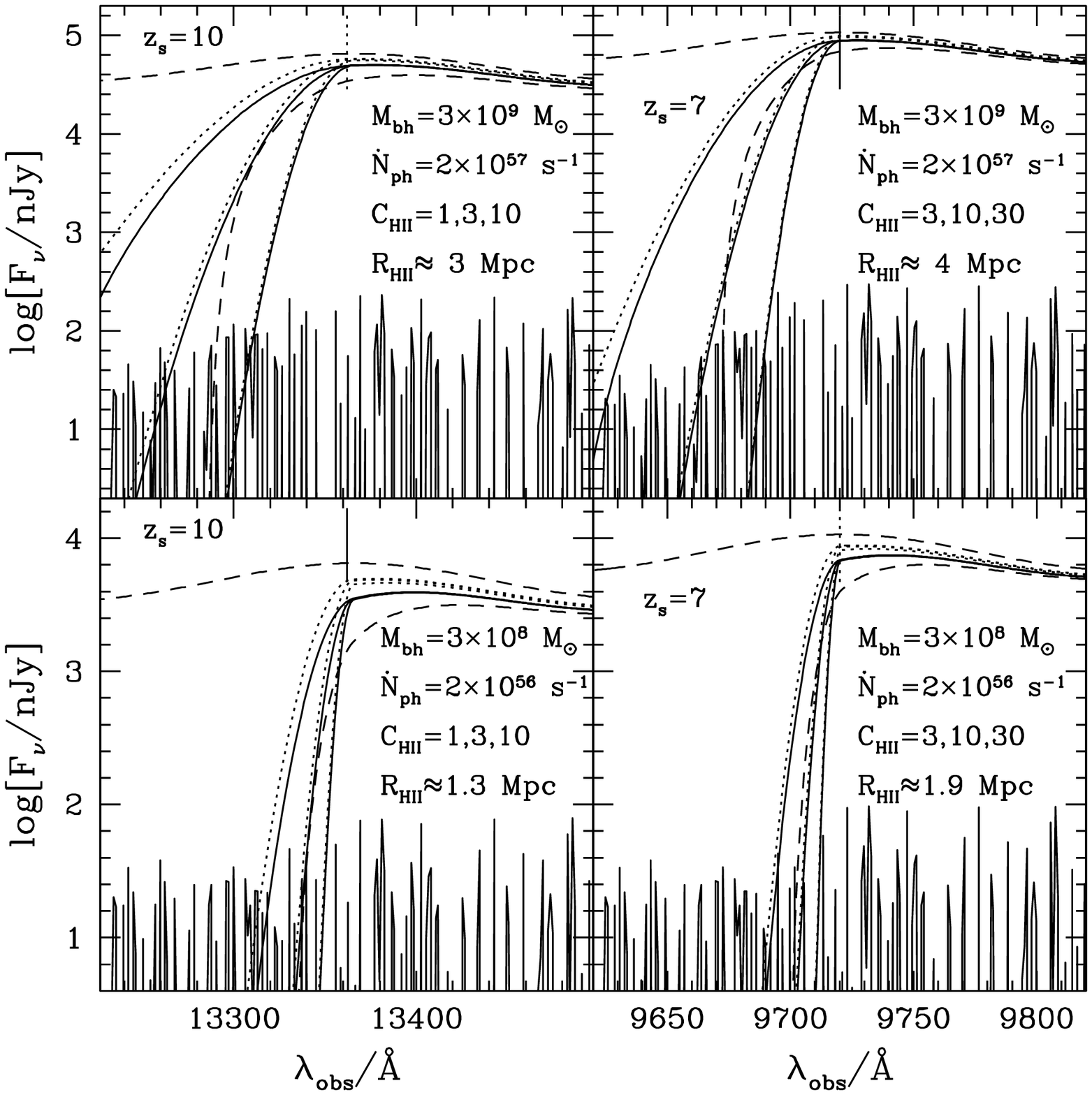}{3.2}{0.45}{-10}{-10} 
\figcaption{The transmitted line profile for the Ly$\alpha$ line of a quasar
with a central black hole mass of either $2\times10^9~{\rm M_\odot}$
(similar to what is inferred for the $z=5.8$ SDSS quasar found by F00,
upper panels), or $2\times10^8~{\rm M_\odot}$ (lower panels).  A
source redshift of either $z_s=10$ (left panels) or $z_s=7$ (right
panels) is assumed.  The upper dashed curve shows the line profile
without any absorption. The set of three solid curves shows the
transmission assuming a quasar lifetime of $10^7$ yr, and using
increasing clumping factors, corresponding (top to bottom) to $C_{\rm
HII}=1,3,10~(z=10)$ and $C_{\rm HII}=3,10,30~(z=7)$.  The dotted lines
show the same results, except for a quasar lifetime of $10^8$ yr. The
lower dashed curve assumes a quasar lifetime of $10^6~$yr, and $C_{\rm
HII}=1$ at $z=10$ and $C_{\rm HII}=1$ at $z=7$. Also shown is the
level of noise expected in a $10^5$ s integration with an 8m
telescope, such as {\it NGST}.
\label{fig:spectra}}
\vspace{\baselineskip}

Let us now translate the theoretical illustration in
Figure~\ref{fig:tau} to observable spectra of quasars around the \lya
emission line shown in Figure~\ref{fig:spectra}.  A logarithmic
ordinate is used to better show the flux over many orders of magnitude
down to the expected noise level (although the spectral line may
appear misleadingly flat). We adopt the black hole (BH) mass and
luminosity inferred from the quasar at $z=5.8$ observed by F00 and
place it either at $z_s=10$ or $z_s=7$ in the top two panels; another
quasar with one tenth of this luminosity is used for the two bottom
panels.  In all cases, the profile of the \lya emission line is
assumed to be Gaussian with a half-width of $1,500$ km/s, and with a
central flux that is twice the continuum.  The random noise shown at
the bottom of each panel in Figure~\ref{fig:spectra} is estimated
assuming a $10^5~$second integration with the {\it NGST} (see
http://augusta.stsci.edu for details on the expected backgrounds) with
a spectral resolution of R=10,000.  Obviously, it is advantageous to
use high resolution spectroscopy to maximize the number of pixels that
probe the structure of the HII region on the blue side of the emission
line (especially for faint and/or very young quasars). At spectral
resolutions this high, the dominant absolute ({\it not relative})
noise is either the Poisson photon noise (for bright sources with
$\dot N_{ph} \gsim 10^{56}~{\rm s}^{-1}$) or the detector noise (for
fainter sources), rather than the zodiacal light or sky background
(see expressions for noise in Gillett \& Mountain 1998). A ground
based telescope such Keck or HET with sufficient spectral resolution
should therefore work just as well as {\it NGST}.

A rich amount of information is contained in the profile of the
line. We see that the red side of the \lya emission line is largely
transmitted even for the fainter quasar adopted at the bottom panels,
consistent with Figure 1 showing an optical depth of $\sim 0.1$ there
(except near the \lya wavelength for the case with $t_Q=10^6~$yr,
where the radius of the Str\"omgren sphere is sufficiently small that
the damping wing of the neutral IGM contributes an optical depth
significantly above unity.)  For the brighter quasar shown in the top
panels the transmission on the red side is insensitive to either $t_Q$
or $C_{\rm HII}$, whereas for the fainter quasar shown in the bottom
panels the transmission on the red side is sensitive to $t_Q$ but not
to $C_{\rm HII}$.  It is therefore possible to measure the lifetime of
a relatively faint quasar (perhaps the majority of the quasars belong
to this category) before reionization is complete.  The primary
uncertain factor is the intrinsic \lya emission profile.

The blue side of the \lya emission line is not completely absorbed.
On the contrary, we see in the top panels of Figure~\ref{fig:spectra}
that for a quasar observed by F00 with $t_{Q}>10^7$ yr there is a
large range $\Delta\lambda\approx 50-150$\AA\, where the flux is at
least ten times above the noise (S/N=10). Even if the quasar is ten
times fainter, one should still be able to detect of order several
tens of spectral pixels if $t_{Q}>10^6$ yr (assuming $R=10^4$).  

More interestingly, the extent of the flux transmission $T_\nu$ on the
blue side is a strong function of $C_{\rm HII}$ but a very weak
function of $t_Q$.  The idealized case in Figure~\ref{fig:spectra}
shows the mean flux processed through a relatively gently fluctuating
medium, with no saturated lines [so that $T_\nu\propto \exp(-\beta_\nu
C_{\rm HII})$, where $\beta|\nu$ is a known constant of order
$0.1-1.0$]. An actual observed flux distribution on the blue side may
contain numerous absorption features due to density fluctuations on
small scales.  In the limit when clumping is caused by highly
overdense regions with a small volume filling factor, the absorption
lines saturate, and the mean transmission could increase, rather than
decrease, for increasing values of $C_{\rm HII}$ (since a larger
fraction of the line of sight probes underdense voids).  In order to
adequately model the absorption profile and to make use of the
additional available information, numerical simulations are needed,
which we will explore in a subsequent paper.  In either case, the blue
side of the \lya line will provide a potentially powerful tool to
directly measure the small scale power at the quasar redshift.

Since the intrinsic \lya emission of a quasar is expected to be
significantly broader than the width of the blue side of the absorbed
spectrum, the latter should not sensitively depend on the exact
intrinsic width of the unabsorbed \lya emission line.  Since the red
side of the profile traces the intrinsic profile well for a bright
quasar, it is possible that fitting the entire observed profile
significantly removes the uncertainty of the unknown intrinsic
profile, as long as it is reasonably symmetric and sufficiently broad.
 
Hogan, Anderson, \& Rugers (1997) and Anderson \etal (1999) have made
the point that the quasar lifetime and ionizing background from the
proximity effect for a HeIII region around the quasar (at lower
redshift) is related and showed that a large HeIII region can be
created by a bright quasar with a reasonable lifetime. They indicated
that for a bright quasar of age $t_Q\sim 10^{7-8}~$yr a large HeIII
region of size $\sim 20$\AA\ can be generated.

Finally, we point out that the existence of a useful HII region around
a quasar requires only that the reionization is not complete.  When
individual HII regions overlap, the universe is not yet completely
reionized, due to the travel delay of ionizing photons from other
sources.  In other words, the emission profile that we show above
should exist even at times significantly later than the overlapping
epoch, but before the average Gunn--Peterson optical depth of the IGM
is reduced to below unity (note that the latter can take a significant
time, see, e.g. Gnedin 2000).  This could potentially significantly
reduce the redshift of quasars that should show useful transmission
profiles.

\section{Conclusions}

A quasar one tenth as luminous as that recently reported by F00, if
placed in the neutral IGM at $z_s\ge 7$ before cosmological
reionization, generates a large HII island (a Str\"omgren sphere).  We
show that for such a quasar a substantial fraction of the flux on both
the blue and red sides of the \lya emission wavelength can be
transmitted.  For the range of wavelengths around the \lya wavelength
where significant transmission of flux occurs, the red damping wing of
the intervening neutral IGM outside the Str\"omgren sphere dominates
the optical depth longward of the \lya wavelength, whereas the
residual neutral hydrogen inside the (highly ionized) Str\"omgren
sphere dominates the optical depth shortward of the \lya wavelength.
While the flux longward of the \lya wavelength is not substantially
suppressed to a large offset in wavelength ($\delta\lambda\ge 100$A),
the blue side of the \lya emission line is more absorbed than the red
side and displays a distinct profile that drops sharply with
decreasing wavelength.  Nevertheless, the flux on the blue side is
clearly detectable even for a quasar 10 times fainter than that found
by F00.  We show that {\it NGST} as well as large ground-based, high
spectral resolution telescopes with a spectral resolution of $R=10^4$
should be able to provide such a detection.

Two applications of such a partially absorbed \lya emission line are
suggested.  First, for a relatively faint quasar (for example, one
tenth as luminous as that found by F00), the red side of the \lya line
profile depends mostly on the age of the quasar and therefore can be
used to infer it.  The main uncertainty is the unknown intrinsic
profile.  Second, for bright quasars, the blue side of the \lya line
profile is also detectable, and depends only very weakly on the
lifetime of a quasar but very strongly on the density fluctuations the
gas around the quasar. Therefore, one can use it to directly measure
the fluctuation of gas on small scales at high redshift.  This may
provide a unique opportunity to probe the small-scale power at very
high redshift, where competing cosmological models differ most.

\acknowledgments
We thank M. Haehnelt for useful comments. This research is supported
in part by the NSF grants AST-9803137 and ASC-9740300, and by NASA
through the Hubble Fellowship grant HF-01119.01-99A, awarded to ZH by
the Space Telescope Science Institute, which is operated by the
Association of Universities for Research in Astronomy, Inc., for NASA
under contract NAS 5-26555.  Conclusions similar to the present paper
were reached independently recently by Madau \& Rees (2000).

\end{document}